\documentclass[twocolumn,prd,aps,superscriptaddress,nofootinbib]{revtex4}
\usepackage{graphicx}
\usepackage{dcolumn}
\usepackage{bm}

\def\x{{\bf x}}
\def\y{{\bf y}}
\def\r{{\bf r}}
\def\k{{\bf k}}

\begin{document}
\voffset = 0.3 true in

\title{Unquenching the Quark Model and Screened Potentials}

\author{Eric S. Swanson\footnote{on leave from the
Department of Physics and Astronomy, University of Pittsburgh,
Pittsburgh PA 15260.}}
\affiliation{
Rudolph Peierls Centre for Theoretical Physics, Oxford University,
Oxford, UK,  OX1 3NP.}

\begin{abstract}
\vskip .3 truecm
The low-lying spectrum of the quark model is shown to be robust under the 
effects of `unquenching'.  In contrast, the use of screened potentials is
shown to be of limited use in models of hadrons. Applications to unquenching
the lattice Wilson loop potential and to glueball mixing in the adiabatic
hybrid spectrum are also presented.
\end{abstract}

\maketitle

\section{Introduction}

Lattice gauge Wilson loop computations convincingly demonstrate that the interquark interaction
grows linearly with distance. However, these computations are, somewhat paradoxically, made in
the pure gauge theory: colour sources and sinks are taken to be static and virtual quarks
are omitted.
It is expected that the introduction of 
light quarks will allow the chromoelectric flux tube to break thereby screening the Wilson loop
confinement potential at a range given by $br = 2M_{qQ} - 2M$ where $b$ is the lattice string 
tension, $M_{qQ}$ is the mass of a light quark -- static source system, and $M$ is the mass of the
heavy quark.

The constituent quark model is similar to quenched lattice gauge theory in the sense that 
nonvalence quark effects are neglected (or are absorbed into parameters) in both formalisms.
Incorporating quark loop (for example, meson creation) effects in the quark model has been a 
longstanding goal of hadronic physics\cite{loop}.  It is therefore natural to 
enquire whether
the use of screened potentials in the constituent quark model can account for some of these
effects. Indeed several papers have appeared recently which use this idea and which claim good
agreement with experiment\cite{Vsc}.

More fundamentally, does the constituent quark model survive unquenching? It is possible that
mass shifts induced by meson loops will severely distort the spectrum. I shall demonstrate
that this need not be the case with the aid of a simple example below.
It is also argued that the use of screened potentials is not a well-founded approach to unquenching
the quark model. 
An obvious problem is that the bound state
spectrum disappears above the `ionization' energy, $2M_{qQ}$. While one may regard this
as satisfactory if hadrons become so broad as to be unrecognizable above the ionization energy,
there is no evidence of this in the spectrum.  Furthermore, the prospect of free quarks emerging
at higher energies is not in keeping with the current understanding of confinement.

A defining characteristic of the quark potential model is that the number of constituents 
in any hadronic system is fixed. How this property is modified in a field theoretic context
can be illustrated by
considering the interaction between electrons in lattice QED. In this case, 
the quenched Wilson loop interaction may be interpreted as a potential
because the imposition of the infinite electron mass limit removes all non-instantaneous interactions
from the theory.  The resulting interaction is then simply $V_{WL}=\alpha/r$. Unquenching QED 
introduces light
fermion loop corrections to the instantaneous interaction (and light-by-light
scattering effects)  and this 
interaction can no longer be interpreted as a potential. 
Specifically, as the momentum transfer increases, the virtual 
fermion pair can go on shell and the number of constituents has changed. Increasing
the momentum transfer further creates highly virtual $e^+e^-$ pairs, and the potential
interpretation of the interaction is regained. 

The generic situation is shown in Fig. 1.
It is typical to focus attention on the lowest adiabatic surface, which has suffered obvious
modification. However, higher surfaces remain and, 
as the dashed line in the figure indicates, the concept of a $q\bar q$ potential is still
possibly useful. This will be demonstrated in the next section.

\begin{figure}[h]
\includegraphics[scale=0.35,angle=-90]{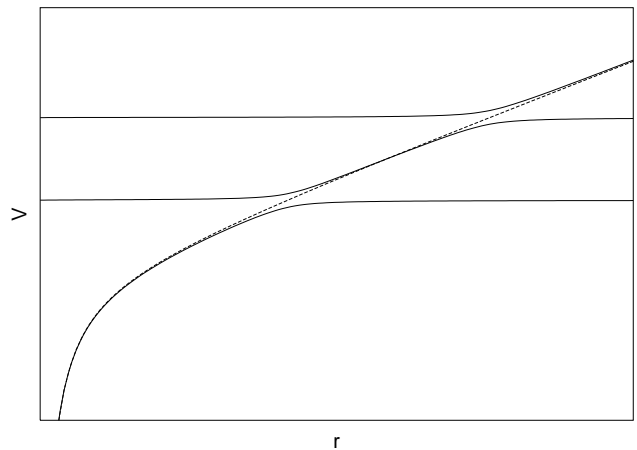}
\caption{\label{3ch} Adiabatic Surface Mixing.}
\end{figure}

\section{A Simple Model of Unquenching}

\subsection{Model Definition}

It is convenient to define a simple field theoretical model which plays the role of full
QCD. The tractability of the model is greatly enhanced by employing
nonrelativistic kinematics. This choice has no bearing on the qualitative
conclusions presented here. The model Hamiltonian is

\begin{eqnarray}
\hat H &=& -\int d^3x \hat\psi_f^\dagger\tau_3 \left(m_f - {\nabla^2\over 2 m_f} \right)\hat\psi_f + g \int d^3x \hat\psi_f^\dagger \tau_1 \hat\psi_f \nonumber \\
&+& {1\over 2} \int d^3x d^3y \hat\psi_{f'}^\dagger(\x)\hat\psi_{f'}(\x) V(\x-\y) \hat\psi_f^\dagger(\y)\hat\psi_f(\y).
\label{qft}
\end{eqnarray}
where the field is defined in terms of particle operators as 

\begin{equation}
\hat\psi_f(\x) = \int {d^3 k\over (2\pi)^{3}} {\rm e}^{i\k\cdot \x}\left(\matrix{b_f^\dagger(\k) \cr d_f(-\k)}\right)
\end{equation}
and the $\tau_i$ are Pauli matrices acting on the quark--antiquark doublet. The flavour index $f$
takes on two values representing a heavy quark species of mass $M$ and a light quark species 
of mass $m$ (denoted $Q$ and $q$ in the following).

The theory is truncated by projecting onto the lowest two Fock space sectors via the
Ansatz

\begin{eqnarray}
|\Psi\rangle &=& \int d^3r_1 d^3r_2 \phi_{QQ}(r_1-r_2)\varphi({r_1+r_2\over 2}) b_Q^\dagger(r_1) d_{Q}^\dagger(r_2) | 0 \rangle  \nonumber \\
&+& \int d^3r_1 d^3r_2 d^3r_3 d^3r_4 \varphi\left({u\over 2}(r_1+r_4) + {v\over 2}(r_2+r_3)\right) \nonumber \\
&& \psi\left( u(r_1-r_4) + v(r_3-r_2)\right) \phi_{Qq}(r_1-r_3) \phi_{qQ}(r_2-r_4)\nonumber \\
&&  b_Q^\dagger(r_1) d_{q}^\dagger(r_3)b_q^\dagger(r_2)d_{Q}^\dagger(r_4)|0\rangle.
\label{Psi}
\end{eqnarray}
Here the wavefunctions $\phi$ represent mesons, $\varphi$ are center-of-mass wavefunctions 
which will be irrelevant in the following, and $\psi$ is the relative two-meson
wavefunction. The parameters $u$ and $v$ are quark mass ratios defined as $u=M/(m+M)$ and
$v = m/(m+M)$. The Ansatz has been specialised to the case representing a $QQ$ meson
and the $Qq\cdot qQ$ continuum.

We shall restrict attention to the light quark loop modifications of heavy
mesons (loop effects on the heavy-light mesons and light-light mesons do not change the
qualitative conclusions presented here).
Computing $\langle \Psi|\hat H | \Psi \rangle$ under these restrictions and varying with 
respect to $\phi$ and $\psi$
yields the coupled equations

\begin{eqnarray}
H_0 \phi_{QQ}(r) + \Omega(r)\psi(ur) &=&  E\phi_{QQ}(r) \nonumber \\
H_1\psi(\rho) + u^{-3} \Omega({\rho\over u}) \phi_{QQ}({\rho\over u}) &=& E\psi(\rho),
\label{fullEq}
\end{eqnarray}
where $r$ is the distance between the quarks in the meson and $\rho$ is the intermeson 
coordinate; these are related by $\rho = ur$.  Additionally,

\begin{equation}
\Omega(\r) = g \int d^3x \phi_{Qq}(\r/2 - \x) \phi_{Qq}(\r/2+\x),
\end{equation}

\begin{equation}
H_0 = 2M - {\nabla^2\over M} + V(r),
\end{equation}
and

\begin{eqnarray}
H_1\psi(\rho) &=& \left[2M_{qQ} - {\nabla^2\over m+M} + V_D(\rho)\right] \psi(\rho) + \nonumber \\
&&  \int d^3R\, V_E(R,\rho)\psi(R) - \nonumber \\
&&  E \int d^3R\, N(R,\rho) \psi(R).
\label{rgm}
\end{eqnarray}
The dependence on $u$ emerges because quark pair creation at a point forces $r_2 = r_3$ and $r_1-r_4 =r$ (see the argument of $\psi$ in Eq. \ref{Psi}). The
factor $u^{-3}$ is due to the integral $\int d^3r \delta(\rho-u r)$  which occurs upon
taking the functional derivative of $\langle \Psi | \hat H|\Psi\rangle$ with respect to $\psi^*(\rho)$. Its presence can be
confirmed by checking hermiticity of Eq. \ref{fullEq}.
Eq. \ref{rgm} describes the interactions of the heavy-light mesons and is in the form of
a resonating group equation\cite{rg}. The kernels $V_E$ and $N$ are nonlocal exchange and 
normalisation kernels whereas $V_D$ is the direct interaction obtained by convoluting the
quark potential with mesonic wavefunctions.
The equations are simplified further by neglecting final state interactions
in the meson-meson channel. Self energy diagrams are assumed to be entirely renormalised away
and a single continuum channel
is used. Again, none of these simplifications qualitatively change the conclusions.
The model specification is complete upon assuming an SHO interaction 

\begin{equation}
V(r) = {1\over 2} k r^2
\end{equation}
with spring constant $k = 0.138$ GeV$^3$ and hence $\omega_{QQ} = 240$ MeV.
This fixes the heavy-light meson wavefunction yielding a transition operator of

\begin{equation}
\Omega(r) = g {\rm e}^{-\beta^2 r^2/4}.
\end{equation}
with $\beta^2 = \omega_{Qq}\,mM/(m+M)$.

Finally quark masses shall be taken to be $M = 4.8$ GeV and $m = 0.3$ GeV. The SHO strength
and the quark masses were chosen so that the model is a facsimile of the upsilon system. As is typical in constituent quark models, a flavour-dependent constant is added to the Hamiltonian. In this case $\delta H_{Qq} = -0.75$ GeV, yielding a `B' meson mass of $M_{qQ} = 5.4$ GeV.

In the following the $q\bar q$ interaction shall be called the `bare' interaction. It is 
denoted
by the solid line in Fig. 2. The screened interaction is obtained from Eq. \ref{fullEq} by
diagonalisation in the static quark limit. The results are the dashed lines in Fig. 2. 
Of course the screened potentials have the expected adiabatic surface crossing at 
$r \approx 4.2$ GeV$^{-1}$. It is also significant that the screened potentials exhibit
 a modified 
short range behaviour. According to lore the adiabatic screened potential is only modified
near the continuum threshold, but, as shown here,  the $r$-dependence of the 
transition amplitude can affect the screened potential at short range. This can have
important phenomenological effects on hadron properties and can, for example, substantially
modify the assumed short range behaviour of the constituent quark model.

\begin{figure}[h]
\includegraphics[scale=0.35,angle=-90]{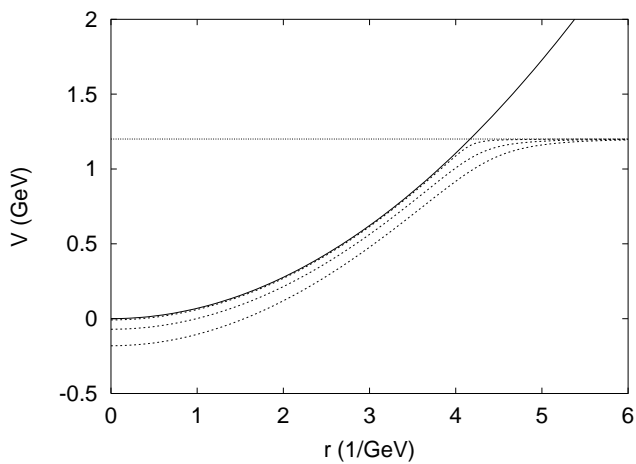}
\caption{\label{Vsc} The Model Screened Potential. The solid line is the
bare potential; the dashed lines are the screened potential for $g=100$, $300$, 
and $500$ MeV from top to bottom.}
\end{figure}

\subsection{Method}

It is convenient to solve the coupled channel problem by integrating out the
bound state channel.  The effective potential
which describes the bound state physics is

\begin{equation}
\langle k| V_{eff} | k' \rangle  = 2\pi^2 \sum_i {\omega_i^*(k)\omega_i(k') \over (E - E_i)}
\end{equation}
where $E_i$ are the eigenenergies of $H_0$ and $\omega_i(k) = \langle \phi_{QQ}^{(i)} | \hat \Omega | k\rangle$.

\begin{figure}[h]
\includegraphics[scale=0.35,angle=-90]{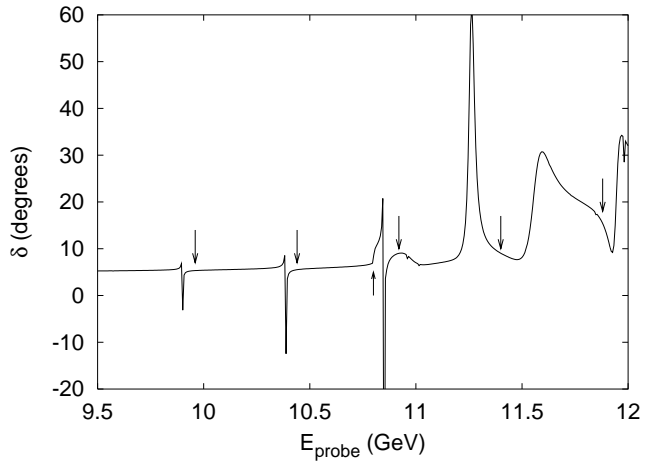}
\caption{\label{MM3a} Probe Phase Shift for $g = 300$ MeV. The arrows indicate
bare pole locations. Threshold is shown at 10.8 GeV.}
\end{figure}

The Bethe-Heitler equation is solved with the addition of a weakly coupled
probe channel to locate the poles in the T-matrix.  A typical computation is
shown in Fig. \ref{MM3a}. Heavy meson poles below threshold are quite narrow
and easily seen by the probe channel. Identifying resonance locations above
threshold is more difficult and hence Argand diagrams were often employed. 
This was essential in the case of the unquenched light meson spectrum because
of the strong scattering which takes place in the $(qq)(qq)$ channel.
In the tables below, 
the real part of the 
poles are referred to as the  `full'  spectrum.

\section{Results and Discussion}

The mass spectra for four possible solutions to the full problem of Eq. \ref{fullEq} are
presented here:

(i) the bare spectrum obtained by solving $H_0 \phi_{QQ} = E\phi_{QQ}$;

(ii) the screened spectrum obtained by employing the screened potential, 
$V_{sc}$ in the bare Hamiltonian;

(iii) the renormalised spectrum;

(iv) the full spectrum obtained by solving the full Hamiltonian, Eq. \ref{fullEq}.

The renormalised Hamiltonian is obtained by fitting the oscillator frequency $\omega_R$ and
quark mass $M_R$ to the full (`experimental') low lying spectrum. This mimics
the procedure followed with the quark model and serves as a point of 
comparison with the full model and the two alternate truncations.

As mentioned above, the use of the
screened potential is frequently adopted in hadronic physics in an attempt to 
model the effects of coupled channels\footnote{Unfortunately, a recent  attempt to use screened potentials to describe heavy mesons\cite{QQsc} is not a good test of the method since the ionization threshold was chosen more
than 1 GeV above the physical threshold and the string tension was adjusted so that the 
screened potential approximated the bare potential in the physically relevant region.}.
The efficacy of this idea will be tested by comparing with the full spectrum below.

\subsection{Heavy Mesons}

Model parameters were chosen to mimic the upsilon system: $M= 4.8$ GeV, $m = 0.3$ GeV and $\omega_{QQ} = 0.24$ GeV. The lightest heavy-heavy meson has a mass of 9.96 GeV, the
lightest heavy-light meson is 5.4 GeV, and the `upsilon' decay threshold lies 
between the second and third heavy-heavy states at 10.8 GeV.

\begin{table}[h]
\caption{Heavy Meson Spectra (GeV) [g=100 MeV]}
\begin{tabular}{l|lllll}
\hline
model & $E_0$ & $E_1$ & $E_2$ & $E_3$ & $E_4$ \\
\hline
bare & 9.960  & 10.440 & 10.920 & 11.400 & 11.880\\
screened & 9.952 & 10.428 & -- & -- & -- \\
renorm & 9.956 & 10.430 & 10.904 & 11.380 & 11.852 \\
full & 9.956 & 10.430 & 10.91 & 11.4 & 11.887 \\
\hline
\end{tabular}
\end{table}

Tables I-III present the four spectra with progressively stronger
decay amplitudes. The case with $g=100$ MeV induces small shifts in the bare spectrum,
as indicated by the small renormalisation: the bare parameters are
$\omega = 240$ MeV and  $M = 4.8 $ GeV, while the renormalised ones are  
$\omega_R = 237$ MeV and $M_R = 4.8$ GeV. The effects on the bare spectrum and on
the renormalisation become progressively larger as the coupling increases. 
For $g = 300$ MeV the renormalised parameters are 
$\omega_R = 243$ MeV and $M_R = 4.767$ GeV. For $g=500$ MeV one obtains
$\omega_R = 252$ MeV and $M_R = 4.713$ GeV.

\begin{table}[h]
\caption{Heavy Meson Spectra (GeV) [g=300 MeV]}
\begin{tabular}{l|lllll}
\hline
model & $E_0$ & $E_1$ & $E_2$ & $E_3$ & $E_4$ \\
\hline
bare & 9.960  & 10.440 & 10.920 & 11.400 & 11.880\\
screened & 9.894 & 10.372 & 10.778 & -- & -- \\
renorm & 9.898 & 10.384 & 10.870 & 11.356 & 11.842 \\
full & 9.898 & 10.384 & 10.847 & 11.255 & 11.578 \\
\hline
\end{tabular}
\end{table}

\begin{table}[h]
\caption{Heavy Meson Spectra (GeV) [g=500 MeV]}
\begin{tabular}{l|lllll}
\hline
model & $E_0$ & $E_1$ & $E_2$ & $E_3$ & $E_4$ \\
\hline
bare & 9.960  & 10.440 & 10.920 & 11.400 & 11.880\\
screened & 9.794 & 10.279 & 10.708 & -- & -- \\
renorm & 9.803 & 10.307 & 10.811 & 11.315 & 11.819 \\
full & 9.803 & 10.307 & 10.758 & 11.29 & 11.63 \\
\hline
\end{tabular}
\end{table}

These results indicate that the screened potential is quite accurate
below the ionisation threshold for heavy quarks. However, it is useless, and in fact, 
unphysical above this point. Alternatively the renormalised spectrum remains useful, 
even above threshold. This is true even for $g=500$ MeV where mass shifts as large as
160 MeV are obtained. For example, upon renormalisation, the error in the binding energy
of the fifth resonance
is  9\%.  Deviations between the full model (`experiment') and the renormalised
model (the `quark model') appear once one has exhausted the available parameters and
appear to get worse as one goes higher in the spectrum.

\subsection{Light Mesons}

One expects the heavy spectrum to be relatively stable under the perturbations caused by 
quark pair creation and this is borne out by the computation above. Alternatively,
light mesons are likely to suffer greater effects. This is investigated here using the
same model as above in the light-light sector. Thus 
$\omega_{qq} = 960$ MeV and the quark 
masses are $M=m=300$ MeV. This
yields a ground state meson of mass 2.04 GeV and a decay threshold of 4.08 GeV.
Table IV shows the resulting spectra for $g=100$ MeV. In this case the 
renormalised parameters are 
$\omega_R = 965$ MeV and $M_R = 286$ MeV. Increasing the coupling to $g=300$ MeV
yields the results of Table V  with parameters
$\omega_R = 958$ MeV and $M_R = 232$ MeV.

\begin{table}[h]
\caption{Light Meson Spectra (GeV) [g=100 MeV]}
\begin{tabular}{l|lllll}
\hline
model & $E_0$ & $E_1$ & $E_2$ & $E_3$ & $E_4$ \\
\hline
bare & 2.04  & 3.96 & 5.88 & 7.80 & 9.72\\
screened & 2.036 & 3.844 & -- & -- & -- \\
renorm & 2.02 & 3.95 & 5.88 & 7.81 & 9.74 \\
full & 2.02 & 3.95 & 5.87 & 7.80 &  9.72 \\
\hline
\end{tabular}
\end{table}

In these cases the screened adiabatic potentials do not perform as well as
in the heavy quark case.
This can perhaps be expected since the screened potential
is based on the adiabatic heavy quark limit of hadronic interactions. 
However, even under the large mass shifts of the 
$g=300$ case, the renormalised model does quite well, with a 1.5\% error in the 
binding energy of the fifth resonance.

\begin{table}[h]
\caption{Light Meson Spectra (GeV) [g=300 MeV]}
\begin{tabular}{l|lllll}
\hline
model & $E_0$ & $E_1$ & $E_2$ & $E_3$ & $E_4$ \\
\hline
bare & 2.04  & 3.96 & 5.88 & 7.80 & 9.72\\
screened & 2.022 & 3.834 & -- & -- & -- \\
renorm & 1.901 & 3.816 & 5.731 & 7.646 & 9.561 \\
full & 1.901 & 3.816 & 5.618 & 7.774 & 9.70 \\
\hline
\end{tabular}
\end{table}

\subsection{Applications to Lattice Computations}

The lessons of the previous section can be profitably applied to lattice gauge 
computations of adiabatic potentials and their mixing. The canonical application
is to the Wilson loop potential.
In particular, extensive effort has been devoted to 
finding `string breaking' in the Wilson loop potential in unquenched QCD\cite{WL}; yet
the expected surface crossing to the meson-meson continuum has been surprisingly difficult
to observe. The common explanation is that better operators are required to see
string breaking. But there is a simpler explanation: the confinement potential and
the meson-meson continuum are connected via a Fock sector transition operator 
which happens to be small. Indeed, the mixed adiabatic surfaces are estimated
to be split by only 50 MeV\cite{GI}. 
Thus surface mixing cannot be observed on the
lattice unless large temporal separations are achieved\cite{GR}. 
A recent lattice computation which finally observes
string breaking confirms these assertions: large temporal extensions in Wilson loops
were required and the calculated
adiabatic potential gap is roughly $50 \pm 4$ MeV\cite{wlb}.

\begin{figure}[h]
\includegraphics[angle=-90,width=8cm]{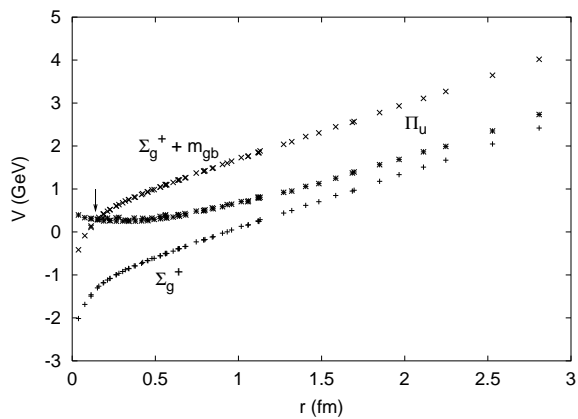}
\caption{Adiabatic Surface Crossing for Heavy Hybrid Potentials. The standard
Wilson loop potential is denoted $\Sigma_g^+$ while the first excited potential
is denoted $\Pi_u$. Lattice data are from Ref.\cite{JKM}.}
\label{vcross}
\end{figure}

A second example from the lattice illustrates the importance and the subtlety of
surface mixing in modelling. We consider the static energy of the first excited 
state of the flux tube in the quenched approximation, denoted as the 
$\Pi_u$ adiabatic potential.
This potential describes the quark dynamics of a heavy hybrid meson state. As seen
in Fig. \ref{vcross}, the $\Pi_u$ potential increases at short distance, in keeping
with the expected perturbative colour octet interaction of $V = + \alpha_s/ (6 r)$.
However, it becomes energetically favourable for the system to emit a gluon 
once the interquark separation becomes sufficiently small.
This gluon combines with the `valence' gluon of the hybrid
to form a scalar glueball, thereby changing the quark colour configuration to that
of a singlet, and hence the adiabatic potential to the attractive $V = - 4 \alpha_s/(3r)$
form. In short, it is energetically favourable for the hybrid to convert to a ground state
meson and a glueball at short distances.
The point at which this should happen is indicated by the arrow in Fig. \ref{vcross}. 
It is thus somewhat perplexing that the lattice data passes through this point
with no transition seen.

It is possible that the expected surface mixing does not occur 
because the minimal relative momentum permitted on the lattices employed in the study 
were too coarse to permit the required hybrid decay. However, a
physical reason for the suppression of this
coupling is also possible. For example,  if one considers the hybrid to be dominated
by a Fock space component consisting of a constituent quark, antiquark, and gluon, then
the postulated transition occurs through gluon emission from either the valence gluon
or a valence quark. In the former case the coupling to a glueball is 
zero due to the colour overlap while in the latter case the coupling is suppressed by
the large (infinite) quark mass. Thus one expects the coupling between the surfaces to
be very small, which implies that the surface mixing will not be seen unless lattices
with exceptionally large temporal extents are employed. 

While this is a plausible explanation for the observed short range behaviour of the
$\Pi_u$ potential it begs the question: which surface should one use when modelling
heavy hybrids? The observations of the previous section dictate that one
should use the unmixed lattice potential -- mixing is a non-potential effect and should
not be considered in valence hybrid models (in the same way that screened potentials 
are not useful for modelling
excited canonical mesons and baryons)\cite{IP}.

\section{Conclusions}

The lessons gleaned from the simple mixing model presented here are
directly applicable to the constituent quark model. One sees, for example, 
that mixing with the continuum can induce an effective short range interaction.
This interaction is in general spin-dependent and can have a substantial effect
on phenomenology. Sorting out spin-dependence due to gluon exchange, instantons,
relativistic dynamics, and continuum mixing remains a serious challenge for
model builders\cite{srs}. Another lesson concerns the lore that passing the lowest
continuum threshold in a given channel is associated with a deterioration 
 of the
quality of potential models of hadrons. But we have seen that hadron masses are
shifted throughout the spectrum (including below threshold) due to a given channel. 
Rather, it is
the proximity of a continuum channel which can cause local distortions of the
spectrum\footnote{Mass shifts due to a single continuum are small for bare
states far from threshold.}. Of course, the problem in hadronic physics is that continuum
channels tend to get dense above threshold.

A number of additional issues complicate the application of these ideas
to hadronic physics.  For example,
the QCD transition operator is determined by nonperturbative gluodynamics and is certainly
not as simple as that of Eq. \ref{qft}.
In addition, as stated above, many continuum channels are present.  Summing over
these is nontrivial -- indeed the sum may not converge (see for example, Ref. \cite{GI}).
Furthermore, one expects that when the continuum virtuality is much
greater than $\Lambda_{QCD}$ quark-hadron duality will be applicable and the sum
over hadronic channels should evolve into perturbative quark loop corrections to
the Wilson loop/quark model potential. Correctly incorporating this into constituent
quark models requires marrying QCD renormalisation with effective models
and is not a simple task. 
Finally, pion and multipion loops can be expected to dominate the virtual continuum
component of hadronic states (where allowed) due to the light pion mass. This raises
the issue of correctly incorporating chiral dynamics into unquenched quark models.
The relationship of chiral symmetry breaking to the constituent quark model has been
discussed in Ref. \cite{ss6} and a variety of hadronic models which incorporate 
chiral symmetry breaking exist\cite{chiral} but much remains to be achieved.

The results presented here imply that, in analogy with the
renormalised model,  constituent quark
models must become progressively less accurate high in the spectrum. However, this 
effect is likely to be overwhelmed by more serious problems: the nonrelativistic 
constituent quark
model must eventually fail as gluonic degrees of freedom are activated and because
chiral symmetry breaking (which is the pedestal upon which the constituent quark
model rests) becomes irrelevant for highly excited states\cite{ess}.
It is clear that an exploration of the excited hadron spectrum is required to
understand the interesting physics behind the unquenched quark model, gluonic
degrees of freedom, and chiral symmetry breaking.

\begin{acknowledgments}

I am grateful to Gunnar Bali, Frank Close, and Chris Michael for discussions and
to Gunnar Bali for  bringing Ref. \cite{GR} to my attention.
This work was supported by PPARC grant PP/B500607 and the US Department of 
Energy under contract DE-FG02-00ER41135.
\end{acknowledgments}


\end{document}